\documentclass[12pt]{article}
\usepackage{graphicx} % standard LaTeX graphics for including eps-figure files
\usepackage{cite}     % this package collapses a list of three or more consecutive reference numbers into a range

\def \d {{\rm d}}
\def \P {P}
\def \e {e}

\begin{document}

\title{\bf Radiative spacetimes approaching the Vaidya metric}

\author{Ji\v{r}\'{i} Podolsk\'{y}  and Otakar Sv\'{\i}tek
\\ \\ \\
Institute of Theoretical Physics, Charles University in Prague,\\
Faculty of Mathematics and Physics, V~Hole\v{s}ovi\v{c}k\'ach 2,\\ 
180~00 Praha 8, Czech Republic }

% 12 April 2005
\date{\today}
\maketitle

%\mailto{podolsky@mbox.troja.mff.cuni.cz}
%\mailto{ota@matfyz.cz}

\begin{abstract}  
We analyze a class of exact type II solutions of the Robinson--Trautman family which 
contain pure radiation and (possibly) a cosmological constant. It is shown that these spacetimes 
exist for any sufficiently smooth initial data, and that they approach the spherically symmetric
Vaidya--(anti-)de~Sitter metric. We also investigate extensions of the metric, and  we 
demonstrate that their order of smoothness is in general only finite. Some applications 
of the results are outlined.
\end{abstract}

\\

\section{Introduction}

The classic Vaidya metric \cite{Vaidya:1943,Vaidya:1951,Vaidya:1953,Stephanietal:book}
(see also \cite{WangWu:1999,Krasinski:1999} followed by reprints of the original Vaidya papers)
is a spherically symmetric type~D solution of the Einstein equations in the presence of pure radiation matter field which propagates at the speed of light. In various contexts this ``null dust'' may be interpreted as high-frequency electromagnetic or gravitational waves, incoherent superposition of aligned waves with random phases and polarisations, or as massless scalar particles or neutrinos. The Vaidya solution 
depends on an arbitrary ``mass function'' ${m(u)}$ of the retarded time $u$ which characterises 
the profile of the pure radiation (it is a
``retarded mass'' measured at conformal infinity). Various sandwiches and shells of
null matter can thus be constructed that are bounded either by flat (${m=0}$) or Schwarzschild-like 
(${m=\hbox{const}\not=0}$) vacuum regions. Due to this property such solutions
have been extensively used as models of spherically symmetric gravitational collapse of a star, 
as an exterior solution describing objects consisting of heat-conducting matter, as an interesting toy model for investigation of singularities and their possible removal by quantum effects, for studies of 
various formulations of the cosmic censorship conjecture on both classical and quantum level, 
process of black-hole evaporation, and for other purposes (see, e.g., \cite{Hiscock:1981a,Hiscock:1981b,HiscockWilliamsEardley:1982,Kuroda:1984a,Kuroda:1984b,BicakKuchar:1997,BicakHajicek:2003,GhoshDadhich:2001,Harko:2003,GirottoSaa:2004}
for more details and related references).

In fact, the Vaidya spacetime belongs to a large Robinson--Trautman class of expanding nontwisting solutions
\cite{RobinsonTrautman:1960,RobinsonTrautman:1962,Stephanietal:book}. 
Various aspects of this family have been studied in the last two decades. 
In particular, the existence,  asymptotic behaviour 
and global structure of \emph{vacuum} Robinson--Trautman spacetimes of type~II with   
spherical topology were investigated
\cite{FosterNewman:1967,Luk,Vandyck:1985,Vandyck:1987,Schm,Ren,Tod:1989,ChowLun:1999,Sin,FrittelliMoreschi:1992},
 most recently in the works of
Chru\'{s}ciel and Singleton \cite{Chru1,Chru2,ChruSin}.
In these rigorous studies, which were based on the analysis 
of solutions to the nonlinear Robinson--Trautman equation for generic, arbitrarily strong
smooth initial data, the spacetimes were shown to exist globally for all positive
retarded times, and to converge asymptotically to a corresponding Schwarzschild
metric. Interestingly, extension  across
the  ``Schwarzschild-like'' event horizon can only be made with a
finite order of smoothness. 
Subsequently, these results were generalized in \cite{podbic95,podbic97} to
the Robinson--Trautman vacuum spacetimes which admit a nonvanishing 
 \emph{cosmological constant} $\Lambda$.
It was demonstrated that these cosmological solutions
settle down exponentially fast to a Schwarzschild--(anti-)de Sitter solution at large
times $u$.  In certain cases the interior of a Schwarzschild--de Sitter black
hole can be joined to an ``external'' cosmological
Robinson--Trautman region across the horizon with a higher
order of smoothness than in the corresponding case with
$\Lambda=0$. For the extreme value $9\Lambda m^2=1$, the extension is smooth but
not analytic (and not unique). The models with  $\Lambda>0$ also exhibit explicitly the
cosmic no-hair conjecture  under the presence of gravitational waves.
On the other hand, when ${\Lambda<0}$ the smoothness of such an
extension is lower.

Our aim here is to further extend the  Chru\'{s}ciel--Singleton 
analysis of the Robinson-Trautman vacuum equation by including matter, namely 
\emph{pure radiation}. It was argued already by Bi\v{c}\'{a}k and Perj\'{e}s \cite{BicakPerjes:1987} 
that with ${\Lambda=0}$ such spacetimes should generically approach the Vaidya metric asymptotically. 
We will analyze this problem in more detail, including also the possibility of
$\Lambda\not=0$ in which case the Robinson--Trautman spacetimes containing 
pure radiation  can be shown to approach the radiating Vaidya--(anti-)de~Sitter metric.

\section{The metric and field equations}
In standard coordinates the Robinson--Trautman metric has the form
\cite{RobinsonTrautman:1962,Stephanietal:book,BicPod99I}
\begin{equation}\label{rob}
\d s^2=-\left(K-2r(\ln{\P})_{,u}-2\frac{m}{r}-\frac{\Lambda}{3}r^2\right)\d u^2-2\d u\d r
      +2\frac{r^{2}}{\P^{2}}\d{\zeta}\d{\bar{\zeta}}\ ,
\end{equation}
where  ${K=\Delta(\ln\P)}$ with ${\Delta\equiv 2\P^{2}\partial_\zeta \partial_{\bar{\zeta}}}$
being the Gaussian curvature of the 2-surfaces ${{2\P^{-2}}\d{\zeta}\d{\bar{\zeta}}}$,  
$\,m(u)$ is the mass function, and $\Lambda$ is the cosmological constant. When  the function ${\P(u,\zeta,\bar{\zeta})}$  satisfies the fourth-order Robinson--Trautman field equation
\begin{equation}\label{feq}
\Delta K+12\,m\,(\ln{\P})_{,u}-4m_{,u}=2\kappa\, n^2\ ,
\end{equation}
the metric describes a spacetime (generally of the Petrov type~II) filled with pure radiation 
field ${T_{\mu\nu}=n^2(u,\zeta,\bar{\zeta})\,r^{-2}\,k_\mu k_\nu}$, where
${\textbf{k}=\partial_r}$ is aligned along the degenerate principal null direction
(we use the convention ${G_{\mu\nu}+\Lambda g_{\mu\nu}=\kappa\, T_{\mu\nu}}$). In particular,
vacuum Robinson--Trautman spacetimes are given by ${n=0}$, in which case $m$
can be set to a constant by a suitable coordinate transformation \cite{Stephanietal:book}.
Vacuum spacetimes~(\ref{rob}) --- possibly with a nonvanishing ${\Lambda}$ ---  thus satisfy
the equation ${12\,m\,(\ln{\P})_{,u}=-\Delta K}$.
These include the spherically symmetric Schwarzschild--(anti-)de~Sitter solution which corresponds
to ${\P_{0}=1+\frac{1}{2}\zeta\bar{\zeta}}$. Indeed, replacing the complex 
stereographic coordinate $\zeta$ with angular coordinates by
${\zeta=\sqrt{2}\,\e^{i\phi}\tan({\theta/2})}$, we obtain
${{2\P^{-2}_0}\d{\zeta}\d{\bar{\zeta}}=\d\theta^2+\sin^2\theta\,\d\phi^2}$,
and ${K_0=\Delta_{0}\ln(\P_{0})=1}$.

Here we will restrict ourselves to nonvacuum cases for which the dependence of the mass function ${m(u)}$ 
on the null coordinate $u$ is only caused by a \emph{homogeneous} pure radiation with the density 
${n^2(u)\,r^{-2}}$. When the mass function $m(u)$ is decreasing, the field equation (\ref{feq}) 
can be naturally split into the following pair, 
\begin{eqnarray}
\Delta K+12\,m(u)\,(\ln{\P})_{,u}&=&0\ , \label{RTequation}\\
\hskip19mm  -2\,m(u)_{,u}&=&\kappa\, n^2(u)\ .\label{RTequation2}
\end{eqnarray}
In fact, it was demonstrated in \cite{BicakPerjes:1987} that such a separation can always be achieved
using the coordinate freedom. It is then possible to reformulate equation (\ref{RTequation}) 
by introducing a $u$-dependent  family of smooth 2-metrics $g_{ab}$ on the submanifold 
${r=\hbox{const}}$, ${u=\hbox{const}}$, such that
${g_{ab}=f(u,\zeta,\bar{\zeta})^{-2}g^{0}_{ab}}$,
where ${g^{0}_{ab}(\zeta,\bar{\zeta})}$  is the metric on a 2-dimensional 
sphere $S^{2}$. Since $g_{ab}$ is of the form  
${2\P^{-2}\d{\zeta}\d \bar{\zeta}}$ in our case, we can write
\begin{equation}\label{PfP0}
\P=f \P_{0}\ ,\quad \P_{0}=1+\textstyle{\frac{1}{2}}\zeta\bar{\zeta}\ ,
\end{equation}
and equation (\ref{RTequation}) becomes
\begin{equation}\label{calabi}
\frac{\partial f}{\partial u}=-\frac{1}{12m(u)}\,f\,\Delta K\ ,
\end{equation}
where $\Delta$ is the Laplace operator associated with the metric $g_{ab}$. Denoting $\Delta_{0}$ and ${K_{0}=1}$ 
as the corresponding quantities related to $g^{0}_{ab}\,$, we obtain
\begin{equation}\label{R+Laplace}
\Delta=f^2\Delta_{0}\ ,\qquad K=f^2(1+\Delta_{0}(\ln f))\ .
\end{equation}

\begin{figure}[t]
\begin{center}
\includegraphics[height=65mm]{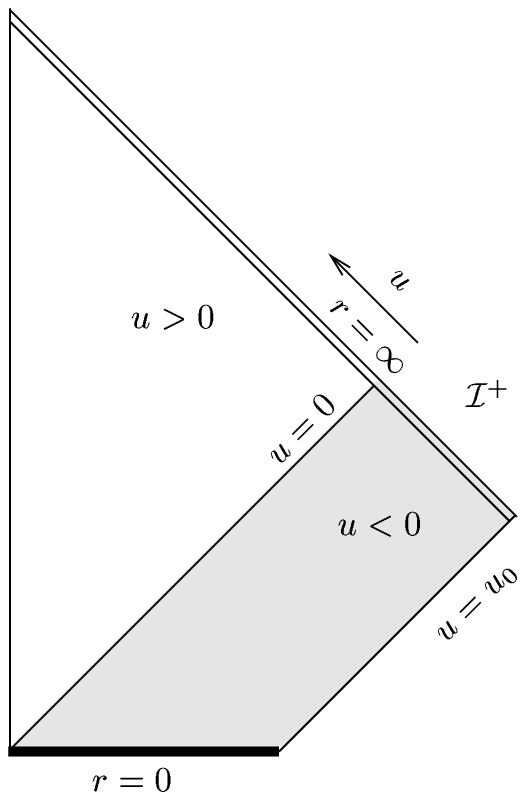}
\includegraphics[height=65mm]{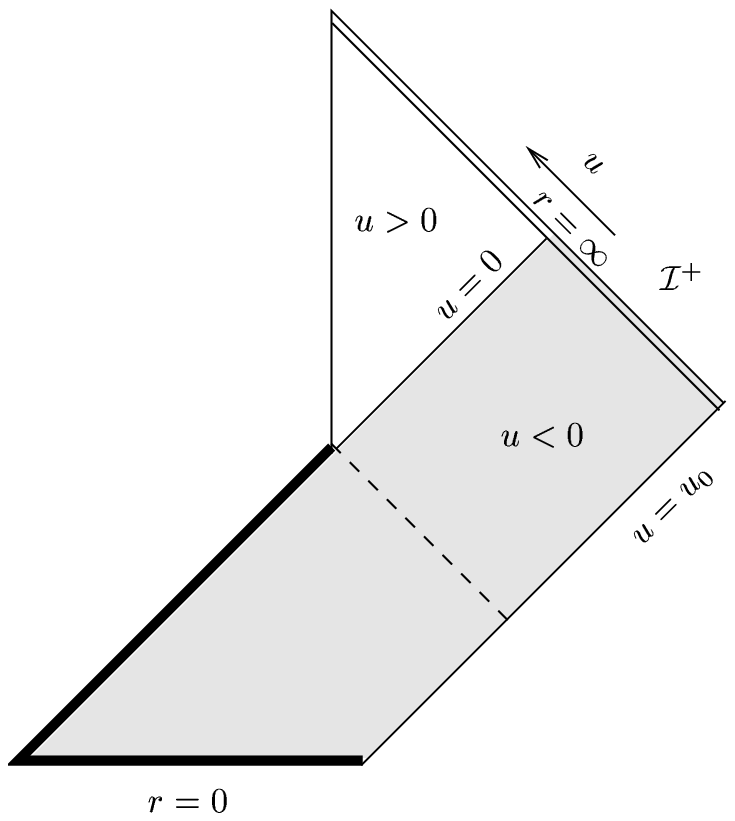}
%\vspace{9pt}
\end{center}
\caption{\label{figure1}%
Schematic conformal diagrams of the Robinson--Trautman exact spacetimes which exist
for any smooth initial data prescribed on ${u_0}$. Pure radiation field is present in the 
shaded region ${u<0}$. Near ${u=0}$ the solutions approach the Vaidya metric, and can be extended to  
flat Minkowski region ${u>0}$. Thick line indicates the curvature singularity at ${r=0}$
whereas double line represents future conformal infinity ${{\cal I}^+}$ at ${r=\infty}$ 
(${\Lambda=0}$ is assumed). The global structure depends on the value of the parameter 
$\mu$ of the linear mass function (\ref{linmass}): left diagram corresponds to
${\mu>1/16}$, the right one applies when ${\mu\leq1/16}$.}
\end{figure}

\section{Linear mass function}
Let us first consider the simplest choice of $m(u)$ which, in fact, has been widely used
in literature (see e.g. \cite{Hiscock:1981a,HiscockWilliamsEardley:1982,WaughLake:1986}): 
we will assume that the mass function is a \emph{linearly} decreasing positive function
\begin{equation}
m(u)=-\mu\,u, \qquad \mu=\hbox{const}>0 \ ,\label{linmass}
\end{equation}
on the interval ${[u_{0},0]}$. Notice that for (\ref{linmass}) the pure radiation field is uniform
because equation (\ref{RTequation2}) implies ${n=\sqrt{2\mu/\kappa}=\hbox{const}}$, independent of 
the retarded time $u$. The constant value ${u_{0}<0}$ localises an initial null hypersurface
(that extends between the curvature singularity at ${r=0}$ and the conformal infinity ${r=\infty}$)
on which an arbitrary sufficiently smooth \emph{initial data} given by the function 
\begin{equation}
f_{0}(\zeta,\bar{\zeta})=f(u=u_{0},\zeta,\bar{\zeta}) \ ,\label{initdat}
\end{equation}
are prescribed, see Fig.~\ref{figure1}.

\subsection{Existence of the solutions}
Now, the idea is to employ the Chru\'{s}ciel--Singleton results \cite{Chru1,Chru2,ChruSin} concerning the analysis 
of the Robinson--Trautman \emph{vacuum} equation, in particular the existence and asymptotic behaviour of its solutions. 
In the vacuum case $m$ in equation (\ref{RTequation}) is constant, and the solution ${f(u,\zeta,\bar{\zeta})}$ of the
characteristic initial value problem (\ref{initdat}) exists and is unique (in spite of the singularity at
${r=0}$). In the presence of pure radiation given by (\ref{linmass}) it is possible to ``eliminate'' the 
variable mass function from the Robinson--Trautman field equation (\ref{calabi}) mathematically 
by a simple reparametrisation
\begin{equation}
\tilde{u}=-\mu^{-1}\ln(-u) \ ,\label{transf}
\end{equation}
cf. \cite{BicakPerjes:1987}. Indeed, equation (\ref{calabi})  is then converted to 
\begin{equation}\label{chrusciel}
\frac{\partial \tilde{f}}{\partial \tilde{u}}=-\frac{1}{12}\,\tilde{f}\,\tilde{\Delta}\tilde{K}\ ,
\end{equation}
where ${\tilde{f}(\tilde{u},\zeta,\bar{\zeta})=f(u(\tilde{u}),\zeta,\bar{\zeta})}$,
${\tilde{K}=\tilde{f}^2(1+\Delta_{0}\ln(\tilde{f}))}$, and ${\tilde{\Delta}=\tilde{f}^2\Delta_{0}}$. 
Notice that the transformation (\ref{transf}) moves the hypersurface ${u=0}$, on which the mass function
$m(u)$ reaches zero, to ${\tilde{u}=+\infty}$.

Chru\'{s}ciel \cite{Chru2} derived the following asymptotic expansion (as ${\tilde{u}\rightarrow \infty}$) for 
the function $\tilde{f}$ satisfying the evolution equation (\ref{chrusciel}) for any smooth initial data ${\tilde{f}_{0}=f_{0}}$ 
on ${\tilde{u}_{0}=-\mu^{-1}\ln(-u_{0})}$, namely
\begin{eqnarray}
\tilde{f}&=&1+f_{1,0}\,\e^{-2\tilde{u}}+f_{2,0}\,\e^{-4\tilde{u}}+\cdots +f_{14,0}\,\e^{-28\tilde{u}}\nonumber\\
&&\qquad+f_{15,1}\,\tilde{u}\,\e^{-30\tilde{u}}+f_{15,0}\,\e^{-30\tilde{u}}+\cdots \label{ftilda}\\
&=&\sum_{i=0}^\infty\sum_{j=0}^{N_i}\,f_{i,j}\,\tilde{u}^{j}\,\e^{-2i\tilde{u}}\ ,\nonumber
\end{eqnarray}
where $f_{i,j}$ are smooth functions on $S^{2}$ such that ${f_{i,j}=0}$ for ${j>0}$, ${i\le14}$. 
The function $\tilde{f}$ thus exists and converges exponentially fast to 1, which means physically that the 
radiative Robinson--Trautman vacuum spacetimes approach asymptotically the Schwarzschild--(anti-)de~Sitter solution 
as ${\tilde{u}\to\infty}$, see relation (\ref{PfP0}). In our case of pure radiation field (\ref{linmass}) 
we  employ the transformation (\ref{transf}) on expression (\ref{ftilda}) to obtain
the following asymptotic expansion of $f$ as ${u\to 0_{-}}$,
\begin{eqnarray}
f&=&1+f_{1,0}\,(-u)^{2/\mu}+f_{2,0}\,(-u)^{4/\mu}+\cdots+f_{14,0}\,(-u)^{28/\mu} \nonumber\\
 &&\qquad-\mu^{-1}f_{15,1}\,\ln(-u)\,(-u)^{30/\mu}+f_{15,0}\,(-u)^{30/\mu}+\cdots \label{fexpansion}\\
&=&\sum_{i=0}^\infty\sum_{j=0}^{N_i}\,f_{i,j}\,[-\mu^{-1}\ln(-u)]^{j}\,(-u)^{2i/\mu}\ .\nonumber
\end{eqnarray}
As a result, for the initial data  (\ref{initdat}) the Robinson--Trautman type~II spacetimes which contain 
uniform pure radiation field with the linear mass function (\ref{linmass}) do exist in the whole 
region ${u_0\le u<0}$.  It is also obvious that the function $f$ approaches $1$  as ${u\to 0_{-}}$ 
(where also ${m(u)\to 0}$) according  to (\ref{fexpansion}). In other words, these spacetimes 
\emph{approach the spherically symmetric Vaidya--(anti-)de~Sitter metric} near ${u=0}$.

The global structure of such spacetimes is schematically indicated on Fig.~\ref{figure1}. In fact,
there are two possibly different conformal diagrams depending on the value of~$\mu$: 
for ${\mu>1/16}$ there is a white hole singularity at ${r=0}$, for ${\mu\leq 1/16}$ 
there is also a naked singularity, see e.g. 
\cite{HiscockWilliamsEardley:1982,WaughLake:1986,GirottoSaa:2004,BicakHajicek:2003} for more details. 
At ${u=0}$ all of the mass ${m(u)}$ is radiated away, and we can attach Minkowski space 
(de~Sitter space when ${\Lambda>0}$,  anti-de~Sitter when ${\Lambda<0}$; the presence of 
the cosmological constant would change the character of conformal infinity ${\cal I}$ which 
would become spacelike or timelike, respectively) in the region ${u>0}$ along the 
hypersurface ${u=0}$. We will now investigate the smoothness of such an extension.

\subsection{Extension of the metric across ${u=0}$ }
It follows from (\ref{fexpansion}) that the smoothness of $f$ on ${u=0}$ is only finite. 
Depending on the value of $\mu$ two different cases have to be discussed separately: 
$2/\mu$ is an integer, and  $2/\mu$ is a real non-integer positive number.

When $2/\mu$ is an integer then due to the presence of the  $\ln(-u)$ term associated with
$f_{15,1}\neq 0$ the function $f$ is of the class $C^{(30/\mu)-1}$. For $\mu$ very small, 
the integer number ${(30/\mu)-1}$ is large so that $f$ becomes  smoothly extendable to 1 
across ${u=0}$ as ${\mu\to0}$. This represents a naked-singularity Robinson--Trautman spacetime
(see the right part of Fig.~\ref{figure1}) unless ${\mu=0}$ which gives flat space 
everywhere. In the limiting case ${\mu=1/16}$ the function $f$ is of the class ${C^{479}}$. 
For the (white hole) Robinson--Trautman spacetimes given by ${\mu>1/16}$ the smoothness is lower. 
However, it is always at least $C^{14}$ because $\mu\leq 2$ in this case.

In the generic case when $2/\mu$ is not an integer the function $f$ is only of the class 
$C^{\{2/\mu\}}$, where the symbol ${\{x\}}$ denotes the largest integer smaller than $x$. 
Again, with ${\mu\to0}$ the function $f$ becomes smoothly extendable. For $\mu<1/16$ the 
function $f$ is at least of the class $C^{32}$, for $\mu>2$ it is not even $C^{1}$ but 
it remains continuous.

To investigate further the smoothness of the metric when approaching the hypersurface ${u=0_{-}}$
which is the analogue of the Schmidt--Tod boundary of vacuum Robinson--Trautman spacetimes
\cite{Tod:1989,Chru2} we should consider the conformal picture using suitable double-null 
coordinates. Such Kruskal-type coordinates for the Vaidya solution with linear mass 
function (\ref{linmass}) were introduced by Hiscock \cite{Hiscock:1981a,Hiscock:1981b,HiscockWilliamsEardley:1982},
see also \cite{WaughLake:1986,GirottoSaa:2004}, 
and we will use this transformation only to replace the coordinate $r$ since the null coordinate 
$u$ is already appropriate. Introducing a new coordinate $w$ by 
\begin{equation}
\d w=\frac{\d u}{u}-\frac{2\d z}{z(2\mu\, z^2-z+2)}\ , \quad\hbox{where}\quad z=-\frac{u}{r}\ ,
\end{equation}
we put the Robinson--Trautman metric with linear mass function into the form
\begin{eqnarray}\label{krusk}
&&\d s^2=-\left(K-1-2\frac{f_{,u}}{f}\,r\right)\d u^2  \nonumber\\
&&\qquad -\left(2r+u+2\mu\frac{u^2}{r}\right)\d u\d w +2\frac{r^{2}}{{\P}^{2}}\d{\zeta}\d{\bar{\zeta}}\ ,
\end{eqnarray}
where ${r(u,w)}$. For the pure Vaidya metric characterized by ${f=1}$ and ${K_0=1}$ the first
term vanishes identically so that the coordinates of (\ref{krusk}) are indeed the
Kruskal-type coordinates for the Vaidya spacetime with a linear mass function.

The smoothness of a general Robinson--Trautman metric (\ref{krusk}) depends only on the smoothness of the metric coefficients $g_{uu}$ and $g_{\zeta \bar{\zeta}}$ (containing the function $f$) since the coefficient $g_{uw}$ tends to $-r$ as ${u\rightarrow 0}$. The smoothness of $g_{\zeta \bar{\zeta}}$ (for any finite~$r$) and of $K$  is the same as of $f$, see (\ref{R+Laplace}). The function $f_{,u}/f$ is evidently one order less smooth than $f$. Consequently, for ${2/\mu}$ being integer or non-integer number, the metric (\ref{krusk}) 
is of the class $C^{(30/\mu)-2}$ or $C^{\{2/\mu\}-1}$, respectively. We again observe that
the spacetimes approaching the linear Vaidya metric with naked singularity (i.e., for small values of the
parameter $\mu$) possess higher order of smoothness at ${u=0}$.

One might be worried about the invariance of our results, namely with respect to a rescaling of the null coordinate ${u(\hat{u})}$ leading to a different smoothness of the function $f$ and of the metric. In order 
to change the smoothness on the hypersurface ${u=0}$ such rescaling must have a singular character there. 
But this would lead to a degeneracy of the metric coefficient $g_{\hat{u}w}$ of the Vaidya metric, which is forbidden. Consequently, the above results are in this sense unique.

We would like to obtain analogous results concerning smoothness of the extension also for a non-zero 
value of the cosmological constant ${\Lambda}$. Unfortunately, as far as we know, there is no \emph{explicit} 
transformation of the Vaidya--de Sitter metric to the Kruskal-type coordinates even for the 
linear mass function (contrary to the Schwarzschild--de~Sitter case \cite{podbic97}). However, it is possible to start with the Vaidya--de Sitter metric
\begin{equation}
\d s^2=-h(u,r)\,\d u^2-2\d u\d r+r^2\d\Omega^2\ ,
\end{equation}
where ${h(u,r)=1+2\mu\,u\,r^{-1}-\frac{\Lambda}{3}r^2}$,
and perform a coordinate transformation
\begin{equation}\label{difer}
\d w=g\,\d u+2\frac{g}{h}\,\d r\ ,
\end{equation}
where $g(u,r)$ is some function. We arrive at the double-null form for the metric
\begin{equation}
\d s^2=-\frac{h}{g}\,\d u\d w+r^2(u,w)\,\d\Omega^2\  .
\end{equation}
Of course, we have to ensure that $\d w$ in (\ref{difer}) is a differential of the coordinate $w$. The integrability condition ($\d^2w=0$) gives the following quasilinear PDE,
\begin{equation}\label{pde}
h^2\frac{\partial{g}}{\partial{r}}-2h\frac{\partial{g}}{\partial{u}}+4\frac{\mu}{r}g=0\ ,
\end{equation}  
for the undetermined function $g$, which is difficult to solve analytically. The method of 
characteristic curves leads to the first-order ODE of the Abel type which has not yet been solved, 
but the existence of its solution is guaranteed. [It is possible to apply the perturbative approach 
starting from the solvable case of the de Sitter metric ($\mu =0$) and then linearise the PDE in 
the parameter $\mu$. The result, however, can not be presented in a useful closed form.] 
For our purposes it suffices to use a general argumentation: the coordinate $u$ is already suitably 
compactified and we are only determining the complementary null coordinate $w$ to obtain 
the Vaidya--de Sitter metric in the Kruskal-type coordinates (which is smooth on ${u=0}$). 
The corresponding Robinson--Trautman metric in these coordinates  differs only by the term 
${g_{uu}(u,r,\zeta,\bar{\zeta})\,\d u^2}$ (which is absent in the Vaidya--de Sitter case in 
the double null coordinates), and by a different metric coefficient 
${g_{\zeta \bar{\zeta}}=r^2f^{-2}P_{0}^{-2}}$, where $r(u,w)$ is finite and smooth when 
approaching the hypersurface $u=0$. The smoothness is thus not affected by the specific transformation (\ref{difer})
 and it is the same as for the vanishing cosmological constant. This is different from 
 vacuum spacetimes with ${m=\hbox{const}\not=0}$ studied in \cite{podbic95,podbic97} because
 in the present case ${m\to0}$ near ${u=0}$, and the influence of~$\Lambda$ on the smoothness
 becomes negligible.

\section{General mass function} 
The results obtained above can be considerably generalized. Inspired by a similar idea outlined 
in \cite{BicakPerjes:1987} we may consider a reparametrisation on the null coordinate $u$  by
\begin{equation}\label{reparam}
\tilde{u}=\gamma(u)\ ,
\end{equation}
where $\gamma$ is an arbitrary continuous strictly monotonous function. We start with the evolution 
equation (\ref{chrusciel}) for which the existence and uniqueness of solutions has been proven, and their 
general asymptotic behaviour (\ref{ftilda}) has been demonstrated. Now,  by applying the 
substitution (\ref{reparam}) in  equation (\ref{chrusciel}) we obtain
\begin{equation}\label{transfevol}
\frac{\partial f}{\partial u}=-\frac{\dot{\gamma}}{12}\,f\,\Delta K\ ,
\end{equation}
(where the dot denotes a differentiation) which is the evolution equation for the 
function ${f(u,\zeta,\bar{\zeta})}$. This is exactly 
the Robinson--Trautman equation (\ref{calabi}) for the mass function
\begin{equation}\label{genermass}
  m(u)=\frac{1}{\dot{\gamma}(u)}\ .
\end{equation}
For a given smooth initial data on ${u_0}$ there thus exists the Robinson--Trautman 
spacetime (\ref{rob}), including the cosmological constant ${\Lambda}$, with the mass function 
(\ref{genermass}). To obtain a positive mass we consider a growing function ${\gamma(u)}$.
Considering (\ref{RTequation2}) this corresponds to a universe filled with  homogeneous pure radiation 
\begin{equation}\label{generradiat}
n^2(u)=\frac{2}{\kappa}\,\frac{\ddot{\gamma}}{\>{\dot{\gamma}}^2}\ .
\end{equation}
For consistency the function $\gamma$ must be convex. An asymptotic behaviour 
of the function $f$ as ${\gamma(u)\to\infty}$ is easily obtained from the
expansion (\ref{ftilda}) by substituting relation (\ref{reparam}).

In particular, the linear mass function (\ref{linmass}) discussed above is a special
case of (\ref{genermass}) for the transformation (\ref{reparam})   of the form (\ref{transf}).
More general explicit solutions can be obtained, e.g., by considering the power function 
\begin{equation}\label{expl}
\gamma(u)=(-u)^{-p}\ , \qquad p>0 \ ,
\end{equation}
which gives 
\begin{equation}\label{expl2}
m(u)=\frac{1}{p}\,(-u)^{1+p}\ , \qquad n^2(u)=\frac{2(p+1)}{\kappa p}\,(-u)^p \ .
\end{equation}
Both functions $m$ and $n$ approach zero as ${u \to0}$. Due to the   
theorems mentioned above, there exist Robinson--Trautman type~II spacetimes 
in the region ${u<0}$ which approach
the spherically symmetric Vaidya--(anti-)de~Sitter metric as ${u\to 0_-}$ with the mass function 
and pure radiation given by (\ref{expl2}). The asymptotic behaviour of such solutions is determined
 by expression (\ref{PfP0}) with
 \begin{equation} \label{fSpec}
f=1+\sum_{i=1}^\infty\sum_{j=0}^{N_i}\,f_{i,j}\,(-u)^{-jp}\,\exp{\left[{-2i(-u)^{-p}}\right]}\ ,\nonumber
\end{equation}
where ${f_{i,j}=0}$ for  ${j>0}$ if ${i\le14}$. Interestingly, the function $f$ is now smooth 
on ${u=0}$ for any power coefficient $p$, but this still does not guarantee that the extension 
into flat region  ${u>0}$ is analytic (see \cite{podbic97} for a similar situation).

Another simple explicit choice is 
\begin{equation}\label{explh}
\gamma(u)=-M^{-1}\ln\left[\sinh(-u)\right]\ , \qquad M>0\ ,
\end{equation}
which implies (see also \cite{GirottoSaa:2004})
\begin{equation}\label{expl2h}
m(u)=M\,\tanh(-u)\ , \qquad n^2(u)=\frac{2M}{\kappa \,\cosh^2u} \ .
\end{equation}
In the region ${u<0}$ the mass function monotonically decreases from $M$ to zero, 
while the pure radiation field grows from zero to  the value ${2M/\kappa}$ as ${u\to 0}$.
Let us note that in this case the integrated radiation density is finite on the interval ${(-\infty,0)}$,
${\int_{-\infty}^{0}n^{2}(u)=2M/\kappa}$. The expansion near ${u=0_-}$ is
 \begin{equation} \label{fSpech}
f=1+\sum_{i=1}^\infty\sum_{j=0}^{N_i}\,f_{i,j}\,(-M^{-1}\ln\left[\sinh(-u)\right])^{j}\,\sinh^{2i/M}(-u)\ .\nonumber
\end{equation}
If ${\,2/M\,}$ is an integer then the function $f$ belongs to the class $C^{(30/M)-1}$, otherwise it
is of the class $C^{\{2/M\}}$.

\begin{figure}[t]
\begin{center}
\includegraphics[height=65mm]{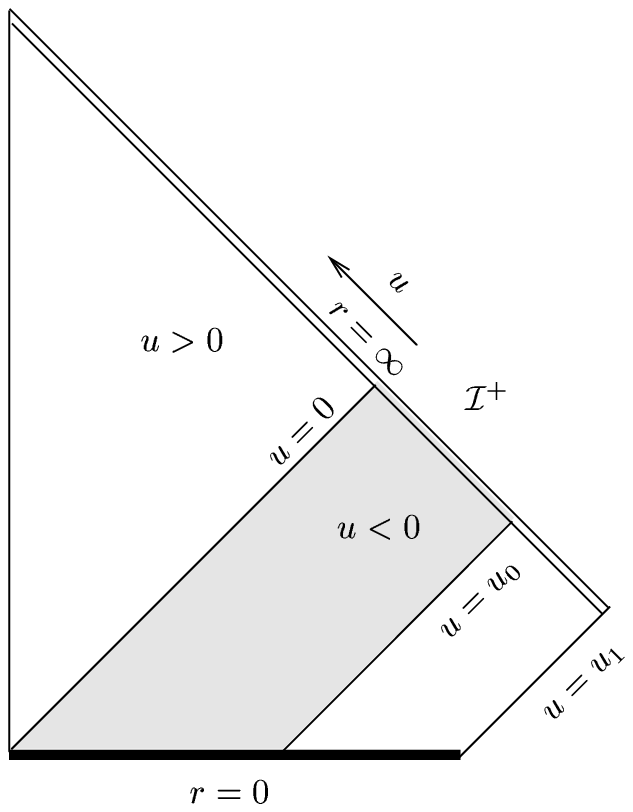}
\includegraphics[height=65mm]{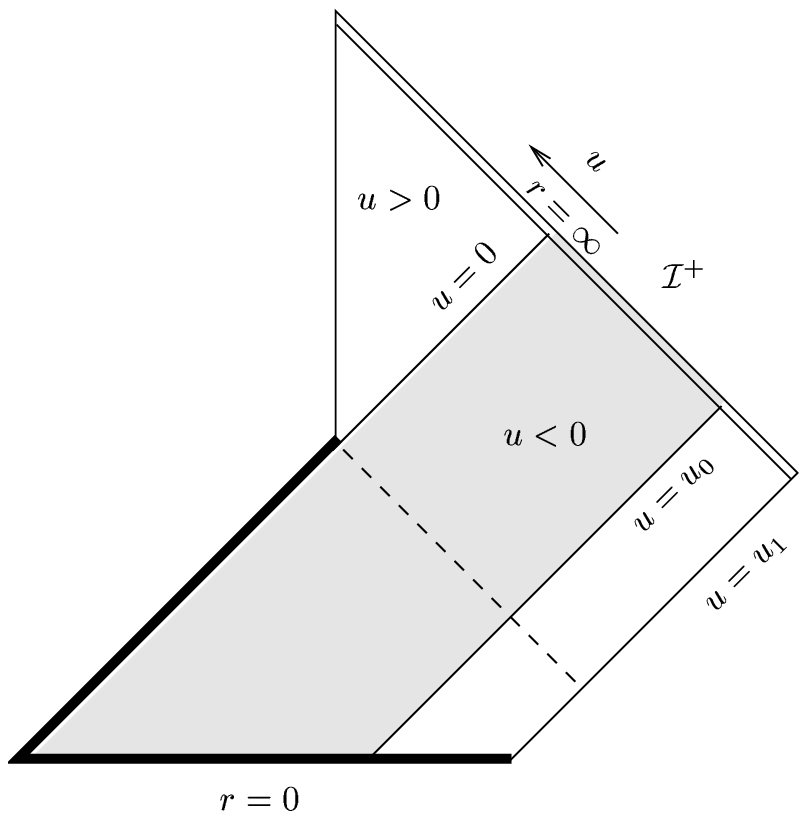}
\end{center}
\caption{\label{figure2}%
Possible extensions of the Robinson--Trautman radiative spacetimes into the region ${u<u_0}$.
Pure radiation is present only in the shaded region, everywhere else it is a vacuum solution.
For ${u\in(u_1,u_0)}$ the mass function is constant, ${m(u_0)=-\mu\, u_0}$, but the spacetime 
is not spherically symmetric --- it is \emph{not} the Schwarzschild solution 
(${\mu>1/16}$ on the left, ${\mu\leq1/16}$ on the right).}

\end{figure}

\section{Possible modifications and applications}

The Robinson--Trautman pure radiation solutions in the region ${u_0\le u\le0}$ approaching
the Vaidya metric near ${u=0}$, which can be extended (albeit non-smoothly) to flat Minkowski 
space in the region ${u\ge0}$ as in Fig.~\ref{figure1},
may be used for construction of various models of radiative spacetimes. For example, it is natural to
further extend the solution ``backwards'' into the region ${u_1< u\le u_0}$ by the Robinson--Trautman 
\emph{vacuum} solution with a constant mass ${m_0=m(u_0)}$, such that the function $f$ is  
continuous on $u_0$. This is shown in Fig.~\ref{figure2}. In such a case the spacetime may
describe the process of ``evaporation'' of a white hole (with a different character of the 
singularity at ${r=0}$ when ${\mu\leq1/16}$) with its mass decreasing from the value 
$m_0$ to zero. Let us emphasize that the region ${u<u_0}$ does not represent the 
Schwarzschild solution because the spacetime is \emph{not spherically symmetric} 
there (${f\not=1}$).  In fact, this is the region where the original 
Chru\'{s}ciel theorems  on the behaviour of the Robinson--Trautman vacuum spacetimes 
with constant mass apply (cf. (\ref{chrusciel}), (\ref{ftilda})). However, the spacetime in this region 
can not be extended up to the past conformal infinity ${\cal I^-}$ because the metric function $f$ diverges as 
${u\to-\infty}$.

In the presence of the cosmological constant $\Lambda$ one obtains a family of exact spacetimes 
that describe evaporation of a white hole in the (anti-)de~Sitter universe. In this case
the schematic conformal diagram on Fig.~\ref{figure2} has to be modified in such a way that for all values of $u$ 
the conformal infinity ${\cal I^+}$ becomes timelike (for ${\Lambda>0}$) or spacelike (for ${\Lambda<0}$).

\begin{figure}[t]
\begin{center}
\includegraphics[height=65mm]{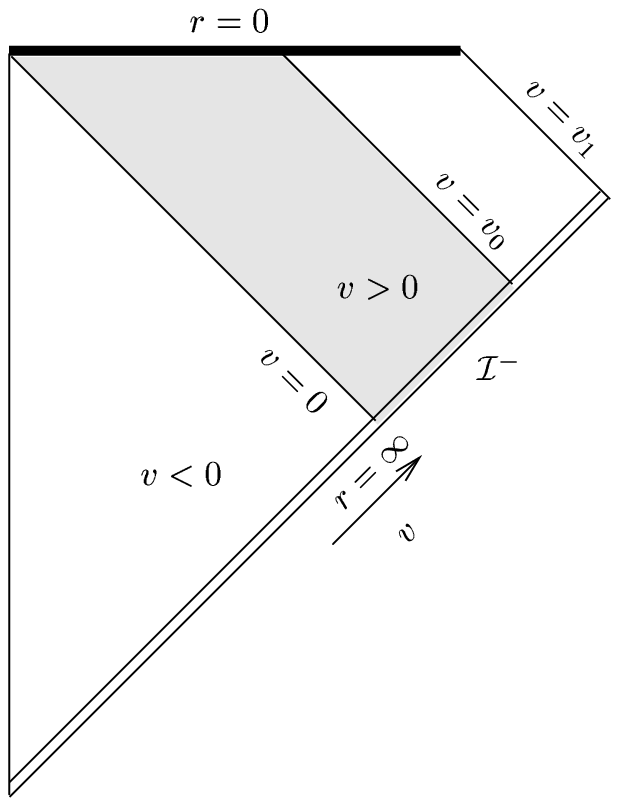}
\includegraphics[height=65mm]{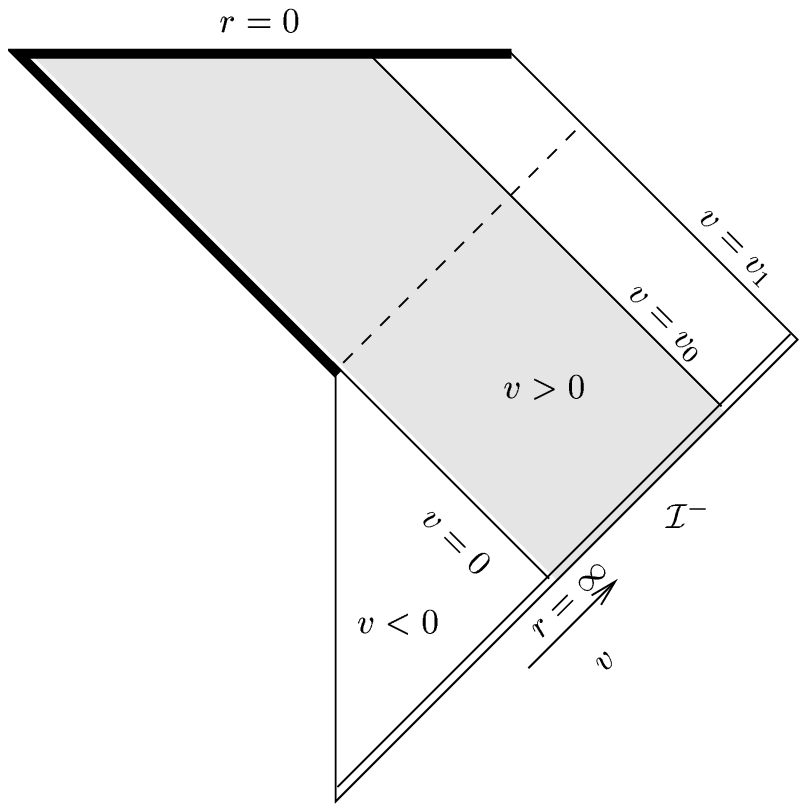}
\end{center}
\caption{\label{figure3}%
Time-reversed version of Fig.~\ref{figure2} represents the ``advanced'' form of the 
Robinson--Trautman spacetimes (\ref{robADV}) which describes an ingoing flow of radiation.}
\end{figure}

Another possible modification is to consider the ``advanced'' form of the spacetimes 
(which describes an ingoing flow) rather than the ``retarded'' form (corresponding to outgoing flow)
employed above (see, e.g., \cite{BicakHajicek:2003} for more details). This time-reversed form is obtained 
formally by a simple  substitution ${u\to -v}$ in the metrics and corresponding functions. 
The Robinson--Trautman metric thus reads
\begin{equation}\label{robADV}
\d s^2=-\left(K+2r(\ln{\P})_{,v}-2\frac{m}{r}-\frac{\Lambda}{3}r^2\right)\d v^2+2\d v\d r
      +2\frac{r^{2}}{\P^{2}}\d{\zeta}\d{\bar{\zeta}}\ ,
\end{equation} 
where $m(v)$ is an increasing mass function in ${v\in[0,v_0]}$. This is joined with flat Minkowskian
region ${v<0}$, and extended to the region ${v\ge v_0}$ by the corresponding 
Robinson--Trautman--(anti-)de~Sitter  black hole vacuum solution, see Fig.~\ref{figure3}. 
It is a non-spherical generalization of the gravitational  collapse of a shell of null dust forming a naked singularity \cite{HiscockWilliamsEardley:1982,DwivediJoshi:1989,DwivediJoshi:1990} ---
in these works the mass function was taken to be ${m(v)=\mu\,v}$  (with ${m(v)=0}$ for ${v\le0}$, 
and ${m(v)=M=\mu\,v_0}$  for ${v\ge v_0}$). The metric function $\P$ is now given by  
${\P=f \P_{0}}$  where $f$ is analogous to (\ref{fexpansion}), 
\begin{equation}\label{fexpansionADV}
f=\sum_{i=0}^\infty\sum_{j=0}^{N_i}\,f_{i,j}\,\left(-\mu^{-1}\ln v\,\right)^{j}\,v^{2i/\mu}\ ,
\end{equation}
so that the smoothness of the metric on the boundary ${v=0}$ depends on the parameter~$\mu$.
For ${v\in(v_0,v_1)}$ the spacetime is vacuum but not spherically symmetric. The metric diverges as 
${v\to\infty}$. Our results can thus be interpreted in such a way that --- at least 
within the Robinson--Trautman family of solutions --- the  
model \cite{HiscockWilliamsEardley:1982} of collapse to a naked shell-focusing singularity 
which is based on the spherically  symmetric Vaidya metric \emph{is not stable} 
against perturbations.

\section{Concluding remarks}
In our contribution we have analyzed  exact solutions of the Robinson--Trautman class which 
contain homogeneous pure radiation and a cosmological constant. This is a natural extension of  previous works 
\cite{FosterNewman:1967,Luk,Vandyck:1985,Vandyck:1987,Schm,Ren,Tod:1989,ChowLun:1999,Sin,FrittelliMoreschi:1992,Chru1,Chru2,ChruSin,podbic95,podbic97,BicakPerjes:1987}
on properties of vacuum spacetimes of this family. We have demonstrated that these solutions 
exist for any smooth initial data, and that they approach the spherically symmetric
Vaidya--{(anti-)}de~Sitter metric. It generalizes previous results according
to which  vacuum Robinson--Trautman spacetimes approach asymptotically the 
spherically symmetric Schwarzschild--(anti-)de~Sitter metric. 
We have investigated extensions of these solutions into Minkowski region, and we 
have shown that its order of smoothness is in general only finite. 
Finally, we suggested some applications of the results. For example, it follows that 
the model of gravitational collapse of a shell of null dust diverges as ${v\to\infty}$ 
which indicates that  investigations of such process  based on the spherically 
symmetric Vaidya metric are, in fact, not stable against ``non-linear perturbations'', 
at least within the Robinson--Trautman family of exact solutions.

\section*{Acknowledgements}
We are grateful to Ji\v r\'\i\  Bi\v c\'ak for valuable comments, and Jerry Griffiths  for
reading the manuscript.

\end{document}